\documentclass[preprint,12pt]{elsarticle}

\usepackage{graphicx}

\usepackage{amssymb,amsmath,color}

\biboptions{numbers,comma,square}

\journal{arXiv}

\DeclareMathOperator{\sech}{sech}

\begin{document}

\begin{frontmatter}



\title{On numerical modeling of dispersive mechanical waves in lipid bi-layers.}


\author[]{Kert Tamm\corref{cor1}}
\ead{kert@ioc.ee}
\cortext[cor1]{Corresponding author}
\author[]{Tanel Peets}
\ead{tanelp@ioc.ee}
\author[]{J\"uri Engelbrecht}
\ead{je@ioc.ee}

\address{Department of Cybernetics, School of Science, Tallinn University of Technology, Akadeemia tee 21, 12618, Tallinn, Estonia}
\begin{abstract}
We investigate different mechanical effects which accompany the nerve pulse propagation by using mathematical modeling. The propagation process is composed by three connected phenomena: (i) the action potential (electrical signal) which is usually considered when nerve pulses are discussed, (ii) the mechanical wave propagating in the biomembrane and (iii) the pressure wave in the axoplasm inside the axon. The main goal of the present study is to investigate numerically how 
the mechanical wave is generated by the action potential and how the characteristics of the system are reflected in the emerging wave ensemble.
The key characteristics for the coupled model system are: (i) the  velocity of the peak of the mechanical pulse is associated with the velocity  of the action potential regardless of the sound velocity value in the lipid bi-layer, (ii) the velocity of the front and the shape of the mechanical wave depends on sound velocity in the lipid bi-layer, (iii) the shape of the mechanical wave can have an effect on the velocity and shape of the action potential. 
\end{abstract}


\end{frontmatter}



\section{Introduction}
\label{sissejuhatus}

We investigate different effects associated with the nerve pulse propagation using mathematical modeling. The propagation process is composed of three connected phenomena: (i) the action potential (electrical signal) \cite{Hodgkin1952,Nagumo1962} which is usually considered when nerve pulses are discussed, (ii) the mechanical wave propagating on the axon surface \cite{Iwasa1980} and (iii) the pressure wave in the axoplasm inside the axon \cite{Terakawa1985}. In this analysis the possible thermodynamical effects \cite{Hady2015,Maugin1994} are left aside. Possibilities for accounting thermodynamic effects within the presented framework can be found in \cite{Tamm2019}. 

While the action potential  of the nerve pulse propagation has been thoroughly studied, the other two effects are not so well understood. For the action potential the Hodgkin--Huxley model \cite{Hodgkin1952} and its modifications are mostly used but there also exists a number of simplified models which are still capable of capturing the key characteristics like, for example, the FitzHugh--Nagumo model (FHN) \cite{Nagumo1962} or the models based on the telegraph equation \cite{Engelbrecht1981,Maugin1994}. In the present paper we use the FHN model for modeling the action potential although it must be stressed that in principle any of the existing action potential models can be used. Here the attention is on coupling effects regarding accompanying phenomena. The main criterion for the used model in the present framework is getting a correctly shaped action potential out of it. 

For the mechanical wave we use the improved Heimburg-Jackson model \cite{EngelbrechtTammPeets2014,Engelbrecht2017,Heimburg2005,Peets2015} which is, in essence, a Boussinesq-type model with nonlinear terms which are of the displacement-type and two dispersive terms (where the second dispersive term is the improvement to the original model \cite{EngelbrechtTammPeets2014}). In the solid mechanics the Boussinesq-type equations are used for investigating nonlinear microstructured materials \cite{Christov2007}. 
The influence of internal structure of solids on wave propagation in such media clearly manifests itself in the wave dispersion. The modeling of such effects can be traced to the Born--von Karman lattice model \cite{BornvonKarman}. Following such an approach, the basic idea is the continualization of discrete systems which brings the higher-order terms in the governing equations into account. An excellent overview on corresponding mathematical models is given by Maugin \cite{Maugin99}. 

For the pressure wave it makes sense to use the classical wave equation as a starting point. Pressure gradients observed experimentally are small \cite{Terakawa1985} allowing one to neglect nonlinear effects. However, axon dimensions are typically in range where viscous effects can be significant enough so the classical wave equation should be improved by taking viscosity into account. Like with other constituents of the present framework the particular model used for describing the phenomenon is not as important as getting a signal shape which is qualitatively similar to experimental observations. Obviously the individual models can be as detailed as needed and what is proposed here should be taken as a starting point and a proof of concept. 

In a nutshell our idea is simple. Take established single models describing all effects related to the nerve pulse propagation that should be considered, propose coupling terms between processes that influence each other, and derive a coupled mathematical model describing an ensemble of waves related to the nerve pulse propagation. As noted, doing all this would be only a first step on a long road ahead as once the single models are collected into one framework, much work remains to be done for the experimental verification and for the proper calibration of the models in such a framework.

\section{Model equations and statement of the problem}
\label{vorrandid}

For modeling the action potential we start with the FHN model \cite{Nagumo1962}
\begin{equation} \label{FHNeq}
\begin{split}
& Z_{T} = Z \left( Z - C_1 - Z^{2} + C_1 Z \right) -I + D Z_{XX},\\
& I_{T} = \varepsilon \left( Z C_2  -I \right),
\end{split}
\end{equation}
where $Z$ is trans-membrane potential, $I$ is the combined ion (recovery) current, $D$ is a coefficient, $C_i = a_i+b_i$ where $a_i$ are the electrical activation coefficients, $b_i$ are the `mechanical' activation coefficients and $X,T$ are the dimensionless spatial and time coordinates, respectively. Subscripts $X,T$ here and further denote partial derivatives with respect to the indicated coordinate. The coefficients $b_i$ could be taken as $b_1 = - g_1 U$ and $b_2 = -g_2  U$ where $g_i$ are constants and $U$ is from Eq.~\eqref{iHJ} -- the feedback from the mechanical wave. Briefly, the underlying idea behind the `mechanical' activation coefficients is that if the density of the lipid bi-layer (biomembrane) increases then it could make it harder for the ions to propagate through the membrane and if it decreases then ions could more easily penetrate through the lipid bi-layer. 

The mechanical wave is modeled by the improved Heimburg - Jackson model \cite{EngelbrechtTammPeets2014,Engelbrecht2017,Heimburg2005}
\begin{equation} \label{iHJ}
\begin{split}
U_{TT} & = c^{2} U_{XX} + N U U_{XX} + M U^{2} U_{XX} + N \left( U_{X} \right)^2 +\\
& + 2 M U \left(U_X \right)^2 - H_1 U_{XXXX} + H_2 U_{XXTT} + g_3 I_X,
\end{split}
\end{equation}
where $U=\Delta \rho$ is the dimensionless longitudinal density change in the lipid bi-layer, $N, M$ are the nonlinear coefficients, $H_1, H_2$ are the dispersion coefficients, $c$ is the velocity in the unperturbed state, $g_3$ is the coupling coefficient and $I_X$ is the gradient of $I$ from Eq.~\eqref{FHNeq}. It should be noted that while recovery current $I$ is a positive pulse-type quantity, its gradient $I_X$ is a bi-polar quantity which is important for the stability of the solution of the Eq.~\eqref{iHJ}. 

\begin{figure}
\includegraphics[width=0.35\textwidth]{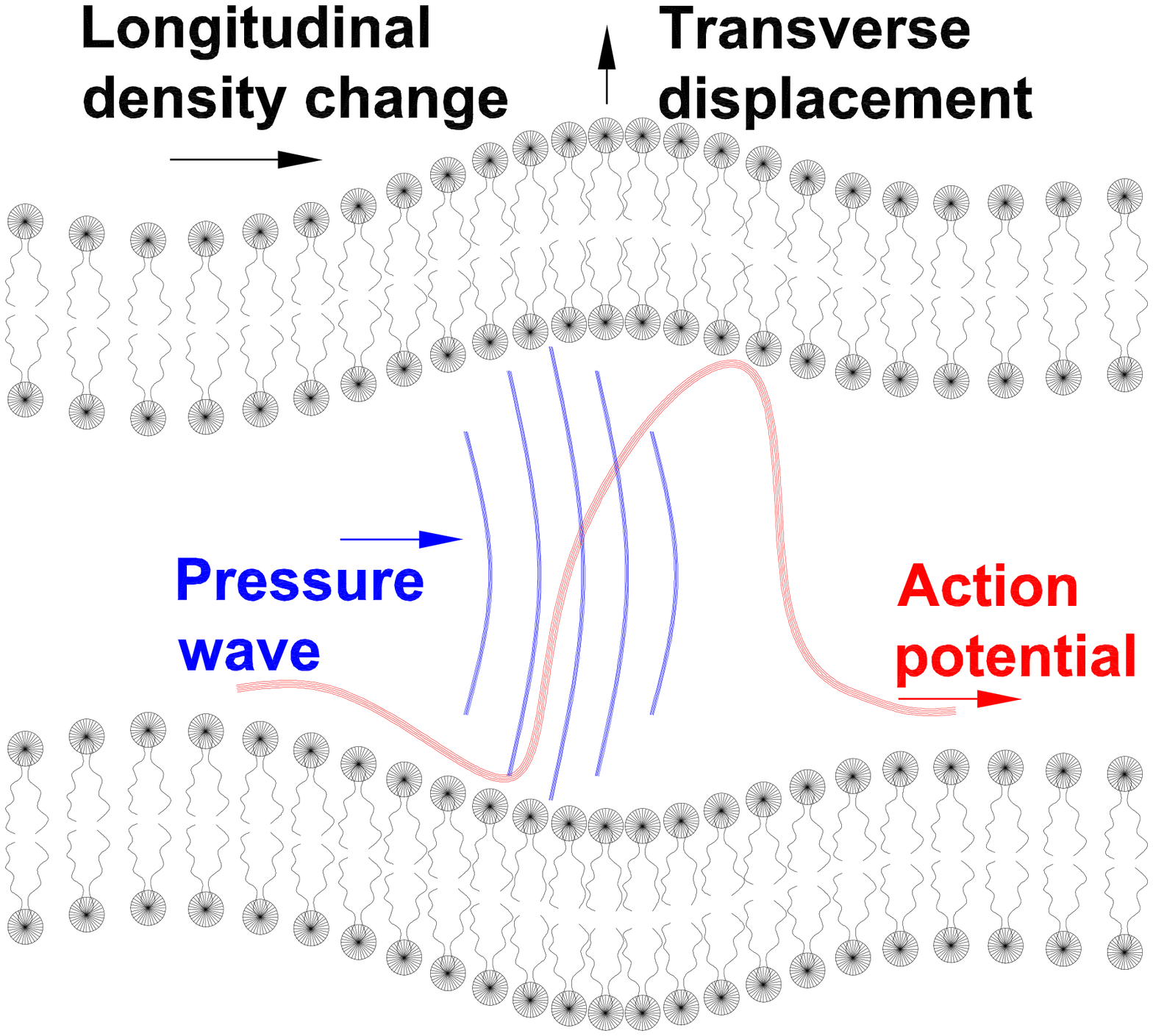}
\includegraphics[width=0.63\textwidth]{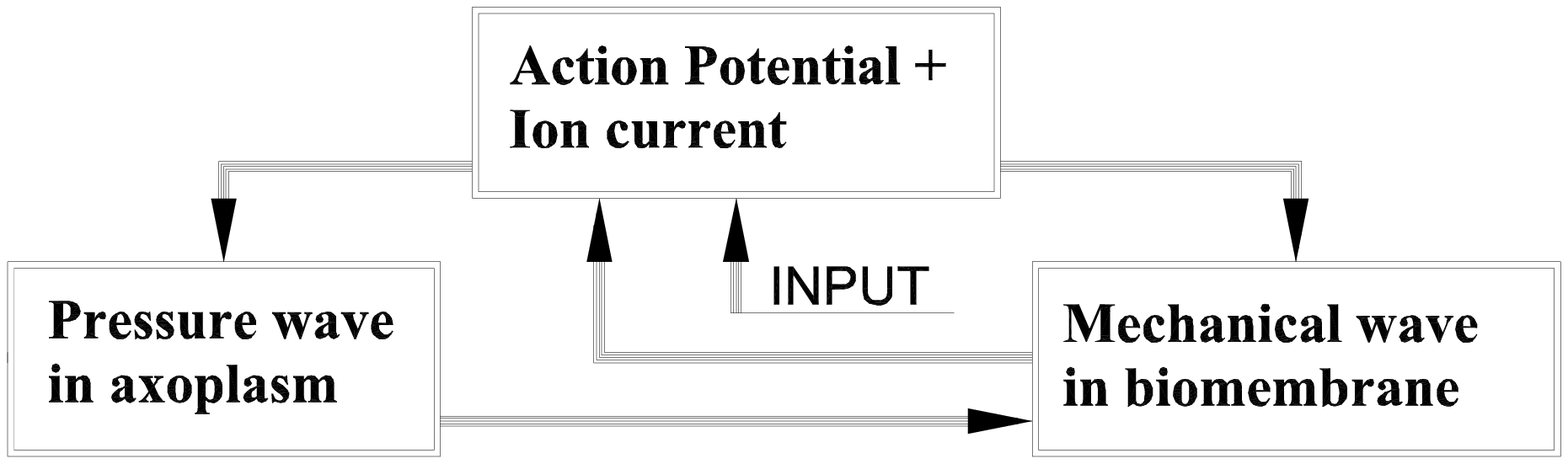}
\caption{Illustrative scheme (right) for the wave ensemble and block diagram of the combined model for the nerve pulse propagation (left).} \label{blokkskeem}
\end{figure}

The coupling term $g_3 I_X$ in Eq.~\eqref{iHJ} can be further elaborated by taking into account the pressure and displacement effects from the ion current(s).
As noted, an improved wave equation equation could be sufficient as a starting point for describing the pressure. As a rough approximation one can assume that the local transverse displacement is proportional to the local pressure change because the transverse displacement observed in the experiments is small \cite{Iwasa1980} compared to the typical diameter of the axon. Here the lipid bi-layer is assumed to be ideally elastic in the presence of such small displacements. 
Then we could use
\begin{equation}\label{Navier1D}
P_{TT} = c_{f}^{2} P_{XX}  - \mu P_T + F(Z,J,U),
\end{equation}
where $P$ is pressure, $c_{f}$ is the sound velocity in axoplasm, $\mu$ is viscosity coefficient. The $F$ is the coupling term accounting for the possible influence from the action potential (ion currents) and mechanical wave in biomembrane. Should the improved wave equation be insufficient it is an option to use some modification, like, for example, the 2D formulation of wave propagation in elastic tube \cite{Lin1956}. The existence and characteristics of a pressure wave accompanying the action potential has been measured by Terakawa \cite{Terakawa1985}.
An illustrative scheme and a block diagram for the proposed combined model are shown in Fig.~\ref{blokkskeem}. 

However, in the present paper a two component model containing the action potential (Eq.~\eqref{FHNeq}) and the mechanical wave (Eq.~\eqref{iHJ}) is used. Simulations including a pressure wave (Eq.~\eqref{Navier1D}) in the numerical scheme can be found in \cite{Engelbrecht2018d}. 

The main goal of the present study is to investigate numerically how 
the mechanical wave is generated by the action potential and how the characteristics of the system are reflected in the emerging wave ensemble.

\section{Initial and boundary conditions, numerical method}
\label{numbrid}

A $\sech^{2}$-type localized initial condition with an initial amplitude $Z_o$ is applied to Eq.~\eqref{FHNeq} and we make use of the periodic boundary conditions
\begin{equation} \label{algtingimus}
Z(X,0) = Z_{o} \sech^2 B_{o} X;
\,
Z(X,T) = Z (X + 2 K m \pi,T);
\,
m = 1,2,\ldots ,
\end{equation}
where $K=160$, i.e., the total  length of the spatial period is $320\pi$.
For Eq.~\eqref{iHJ} we take initial excitation to be zero and make use of the periodic boundary conditions. The solution representing the mechanical wave described by Eq.~\eqref{iHJ} is generated over time as a result of coupling between the Eqs \eqref{FHNeq} and \eqref{iHJ}.

For numerical solving of the Eqs~\eqref{FHNeq} and \eqref{iHJ} we use the discrete Fourier transform (DFT) based pseudospectral method (PSM) \cite{Fornberg1998,salup2009}. 
Variable $Z$ is represented in the Fourier space as
\begin{equation} \label{dft}
\widehat{Z}(k,T) = \mathrm{F} \left[ Z \right]= \sum^{n-1}_{j=0}{Z(j \Delta X, T) \exp{\left(-\frac{2 \pi \mathrm{i} j k}{n} \right)}},
\end{equation}
where $n$ is the number of space-grid points ($n=2^{12}$ in the present paper), $\Delta X=2 \pi/n$ is the space step, $k=0,\pm1,\pm2,\ldots,\pm(n/2-1),-n/2$; $\mathrm{i}$ is the imaginary unit, $\mathrm{F}$ denotes the DFT and $\mathrm{F}^{-1}$ denotes the inverse DFT. 
Basically, the idea of the PSM is to approximate space derivatives by making use of the DFT 
\begin{equation} \label{dft2}
\frac{\partial^{m} Z}{\partial X^{m}} = \mathrm{F}^{-1}\left[(\mathrm{i} k)^{m} \mathrm{F}(Z) \right],
\end{equation}
reducing therefore the partial differential equation (PDE) to an ordinary differential equation (ODE) and then to use standard ODE solvers for integration with respect to time. 

The regular PSM algorithm is derived for $ u_t = \Phi(u,u_x, u_{2x},\ldots,u_{mx})$ type equations. In our case, however, we have a mixed partial derivative term $H_2 U_{XXTT}$ in the Eq.~\eqref{iHJ} and thus the standard PSM has to be modified \cite{lauriandrus2009,lauriandruspearu2007,salup2009}. 
Therefore we rewrite the Eq.~\eqref{iHJ} so that all partial derivatives with respect to time are at the left-hand side of the equation 
\begin{equation} \label{LHSofHE}
\begin{split}
U_{TT} - H_2 U_{XXTT} & = c^{2} U_{XX} + N U U_{XX} + M U^{2} U_{XX} + N \left( U_{X} \right)^2 +\\
& + 2 M U \left(U_X \right)^2 - H_1 U_{XXXX}  + g_3 I_X
\end{split}
\end{equation}
and introduce a new variable
$\Phi = U - H_2 U_{XX}.$
After that, making use of properties of the DFT, one can express the variable $U$ and its spatial derivatives in terms of the new variable $\Phi$:
\begin{equation}\label{UUXPhi}
U=\mathrm{F}^{-1}\left[\frac{\mathrm{F}(\Phi)}{1+H_2 k^2}\right],
\,
\frac{\partial^m U}{\partial X^m}
=\mathrm{F}^{-1}\left[\frac{(\mathrm{i} k)^m \mathrm{F}(\Phi)}{1+H_2 k^2}\right].
\end{equation}
Finally, Eq.~\eqref{iHJ} can be rewritten in terms of the variable $\Phi$ as 
\begin{equation} \label{HEtegelik}
\begin{split}
\Phi_{TT} & =
c^2 U_{XX} + N U U_{XX} + M U^{2} U_{XX} + N \left( U_{X} \right)^2 + 2 M U \left(U_X \right)^2 - \\
& - H_1 U_{XXXX}  + g_3 I_X. 
\end{split}
\end{equation}
In Eq.~\eqref{HEtegelik} all partial derivatives of $U$ with respect to $X$ are calculated in terms of $\Phi$ by using  expression \eqref{UUXPhi} and therefore one can apply the PSM for numerical integration of Eq.~\eqref{HEtegelik}. The FHN model \eqref{FHNeq} is already written in the form of two first-order PDE's which can be solved by the standard PSM without any further modifications.

The calculations are carried out with the Python package SciPy \cite{SciPy}, using the FFTW library \cite{FFTW3} for the DFT and the F2PY \cite{F2PY} generated Python interface to the ODEPACK FORTRAN code \cite{ODE} for the ODE solver. We used `vode' with options nsteps$=10^7$, rtol$=10^{-10}$, atol$=10^{-12}$ and $\Delta T = 0.1$.

\section{Model parameters}
\label{parameetrid}

We solve Eqs \eqref{FHNeq} and \eqref{iHJ} under localized initial conditions and periodic boundary conditions \eqref{algtingimus}. The parameters for the model equations are $D=1$, $\varepsilon = 0.01$ or $\varepsilon = 0.05$, $a_1=0.2$, $a_2=0.2$, $N=0.05$, $M=0.02$, $H_1=0.5$, $H_2=0.75$, $g_1=0.05$, $g_2=0.05$, $g_3=0.02$. Note that as the model Eqs \eqref{FHNeq} and \eqref{iHJ} are dimensionless then also the model parameters are here dimensionless. The parameter relations to the quantities with dimensions can be found, for example, in \cite{EngelbrechtTammPeets2014}. As noted, the length of the spatial period is $320\pi$ and the localized initial condition is given in the middle of the spatial period in the form of sech$^2$-type pulse. The initial condition parameters are $Z_0 = 2$ and the pulse width parameter $B_0 = 1$. It should be noted that if the time-scale parameter is taken as $\varepsilon = 0.01$ then such an initial condition is above the threshold leading to emergence of the propagating AP. However, if $\varepsilon = 0.05$ then the initial pulse is not sufficient for generating the propagating AP.  Such an initial condition is, in essence, a very narrow high amplitude `spark' for the FHN model \eqref{FHNeq}. If the `spark' is above the threshold then the normal FHN pulse is generated in the middle of the spatial period which starts to propagate from there to the positive and negative coordinate $X$ directions. Due to the coupling between the FHN model \eqref{FHNeq} and the improved Heimburg-Jackson model \eqref{iHJ} the mechanical wave is generated in the biomembrane. The initial condition for the Eq.~\eqref{iHJ} is that the system is at rest (zero initial conditions). If the initial condition for the Eq.~\eqref{FHNeq} is below the threshold then the initial electrical signal vanishes over a short time but a small amplitude mechanical wave is still generated as a function of recovery current gradient until the initial excitation vanishes.   

It should be noted that the improved Heimburg-Jackson equation is conservative, i.e., it does not generate nor lose energy, however, due to the added coupling with the FHN equation \eqref{FHNeq} with the improved Heimburg-Jackson equation it is no longer conservative as it can change energy with the Eq.~\eqref{FHNeq}. 

\begin{figure}[h]
\includegraphics[width=0.99\textwidth]{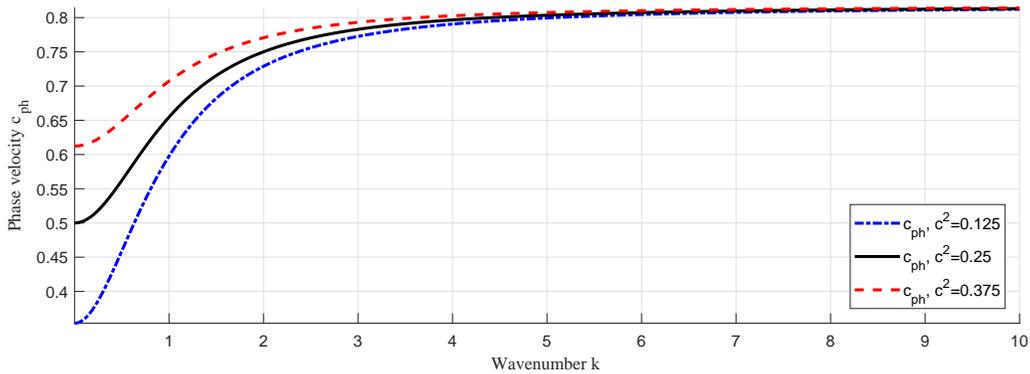}
\caption{Phase velocity graphs for Eq.~\eqref{iHJ} under different $c^2$ values. Dispersion parameters $H_1=0.5$, $H_2=0.75$.} \label{phase}
\end{figure}

The improved Heimburg-Jackson model has two dispersive terms controlled by dispersion parameters $H_1$ and $H_2$. The dispersion relation for Eq.~\eqref{iHJ} can be written as
\begin{equation}\label{dispersioon}
\omega = \sqrt{\frac{c^2 k^2 + H_1 k^4}{1+ H_2 k^2}},
\end{equation}
where $k$ is the wave number and $\omega$ is the dimensionless frequency. 
The ratio of dispersion parameters $H_{1}^{2}/H_{2}^{2}$ determines the bounding velocity in the system while parameter $H_2$  determines the rate at which the bounding velocity is achieved as the frequency of the signal increases. 
If we have the normal dispersion ($c^{2} > H_{1}^{2}/H_{2}^{2}$) then this means that group speed $c_{gr}$ (which usually is the speed of energy propagation in the system) is smaller than the phase speed $c_{ph}$ and if we have an anomalous dispersion ($c^{2} < H_{1}^{2}/H_{2}^{2}$) then it is the opposite, i.e., $c_{gr} > c_{ph}$ (for more details see \cite{Brillouin,Brillouin1960,EngelbrechtTammPeets2014,Engelbrecht2017,Peets2015}). In essence the second dispersive term $U_{XXTT}$ removes the ability of sufficiently high frequencies to travel at near infinite velocities (i.e., makes the model causal). 
Since the mixed derivative term $U_{XXTT}$ reflects the inertial properties then the parameter $H_2$ is related to the inertial properties of the biomembrane \cite{Peets2015}.
Based on experimental considerations \cite{Heimburg2005} the dispersion type for the lipid bi-layer should be anomalous which is the case under the considered model parameters. The phase velocity curves for the Eq.~\eqref{iHJ} under the considered parameters are shown in Fig.~\ref{phase}.


\section{Results and discussion}
\label{tulemused}
\begin{figure}[h]
\includegraphics[width=0.99\textwidth]{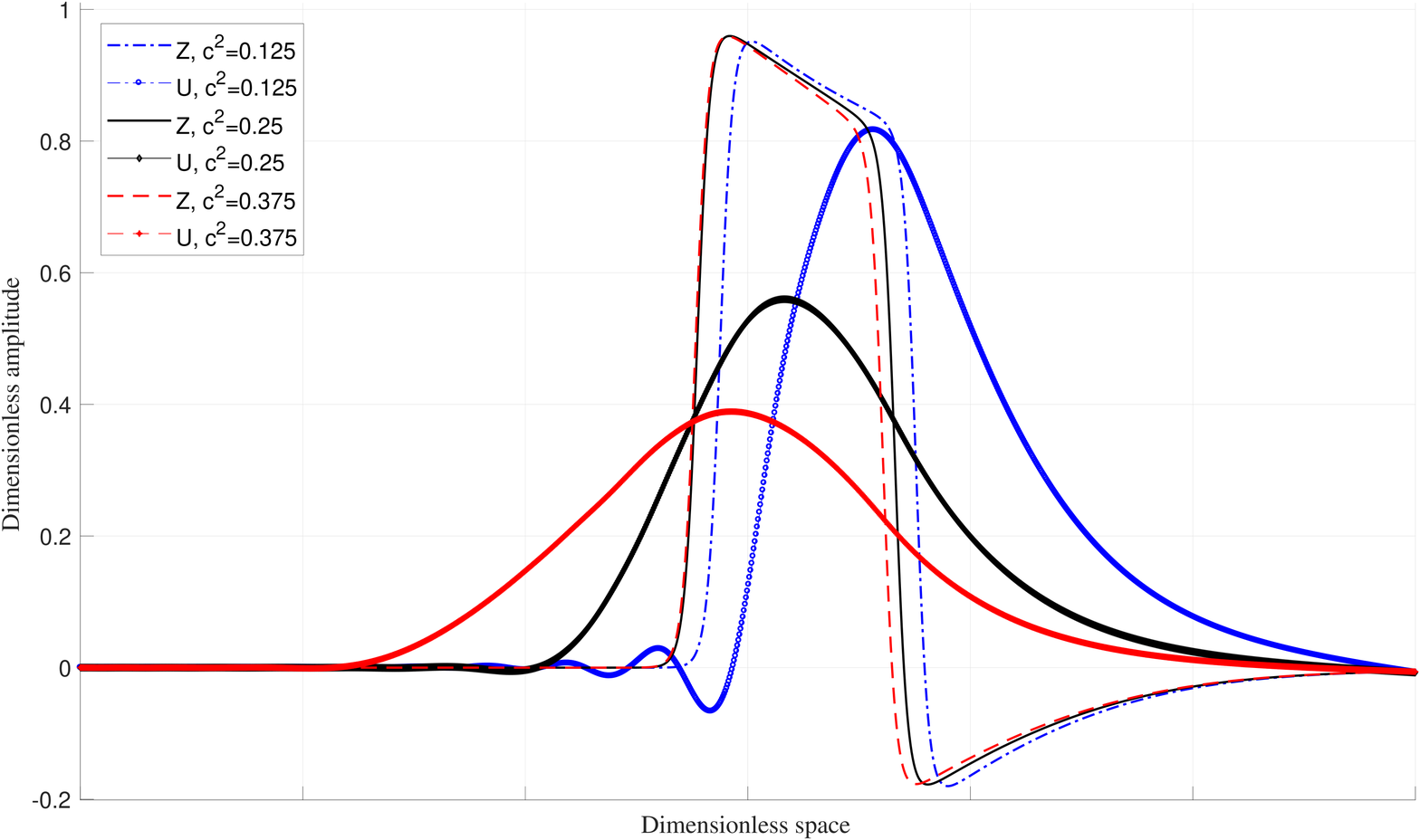}
\caption{The left propagating solutions of Eqs.~\eqref{FHNeq} and \eqref{iHJ}. Parameters $D=1$, $\varepsilon = 0.01$, $a_1=0.2$, $a_2=0.2$, $N=0.05$, $M=0.02$, $H_1=0.5$, $H_2=0.75$, $g_1=0.05$, $g_2=0.05$, $g_3=0.02$.} \label{Fig1}
\end{figure}
\begin{figure}[h] 
\includegraphics[width=0.99\textwidth]{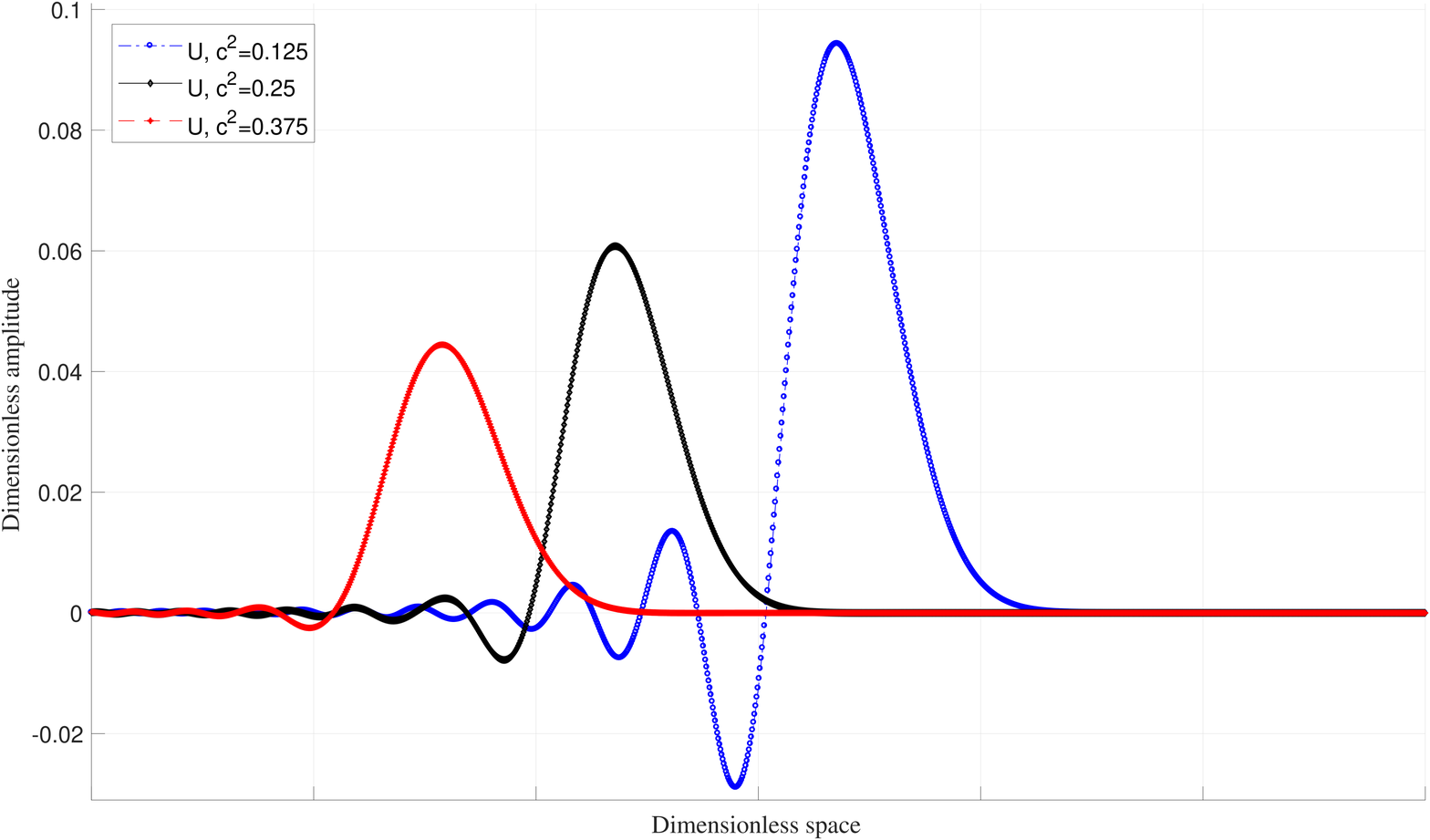} 
\caption{The left propagating solutions of Eq.~\eqref{iHJ}. Parameters $D=1$, $\varepsilon = 0.05$, $a_1=0.2$, $a_2=0.2$, $N=0.05$, $M=0.02$, $H_1=0.5$, $H_2=0.75$, $g_1=0.05$, $g_2=0.05$, $g_3=0.02$.} \label{Fig2}
\end{figure}

The solutions for the coupled model equations \eqref{FHNeq} and \eqref{iHJ} at $T=400$ are shown in Figs~\ref{Fig1} and \ref{Fig2}. As noted previously, the initial `spark' either leads to a formation of the propagating FHN action potentials which then leads to  formation of mechanical waves through coupling (Fig.~\ref{Fig1}) or if the initial value of the `spark' is below the threshold for the FHN model \eqref{FHNeq} then it vanishes over a short time leading to a formation of only a small amplitude mechanical wave which proceeds to propagate without the FHN action potential (Fig.~\ref{Fig2}). Note the difference in the scale of Figs~\ref{Fig1} and \ref{Fig2}. Here only the waves propagating to the left are shown. There exist similar wave-profiles propagating to the right which are not shown. 

In Fig.~\ref{Fig1} we compare action potentials $Z$ and mechanical waves $U$ at dimensionless time $T=400$ provided the velocity of the low frequency sound is different for the lipid bi-layer. A number of essential characteristics for the coupled model system can be observed: (i) the  velocity of the peak of the mechanical pulse is similar to the velocity  of the action potential regardless of the sound velocity value in the lipid bi-layer within the considered parameter range, (ii) the velocity of the front and the shape of the mechanical wave depends on sound velocity in the lipid bi-layer for the Eq.~\eqref{iHJ}, (iii) the shape of the mechanical wave can have an effect on the velocity and shape of the action potential (if $c^2=0.25$ for the mechanical wave then the corresponding action potential $Z$ has propagated further compared with the case $c^2=0.125$). The mechanisms for changing the sound velocity in the lipid bi-layer and different scenarios depending on the mechanical disturbance velocity when compared against the velocity of an electrical pulse have also been studied by El Hady and Machta using different theoretical considerations \cite{Hady2015}.

In Fig.~\ref{Fig2} one can see the wave-profiles of the mechanical wave at $T=400$ when the initial excitation for the Eq.~\eqref{FHNeq} is below the threshold at three different values for the $c^2$. It can be noted that while the front of the wave-profile propagates at the same velocities as in Fig.~\ref{Fig1}, the peak of the packet travels faster than the corresponding wave-profile in Fig.~\ref{Fig1}. While not shown here it should be noted that the system can support also  solitonic solutions, for example, should we take  $c^2=0.755$ under the considered parameters (when $\varepsilon=0.05$) we would get a solution where the initial pulse is separated into a train of solitonic pulses while in the shown cases we have an oscillatory structure instead of the solitonic pulses. The solitonic solutions for the Eq.~\eqref{iHJ} have been studied previously in \cite{Engelbrecht2017,Peets2015,PeetsTammEngelbrecht2017} and for the original model with the simplified dispersion in \cite{Engelbrecht2017,Heimburg2005,Lautrup2011,Vargas2011}.

\begin{figure}[h]
\includegraphics[width=0.46\textwidth]{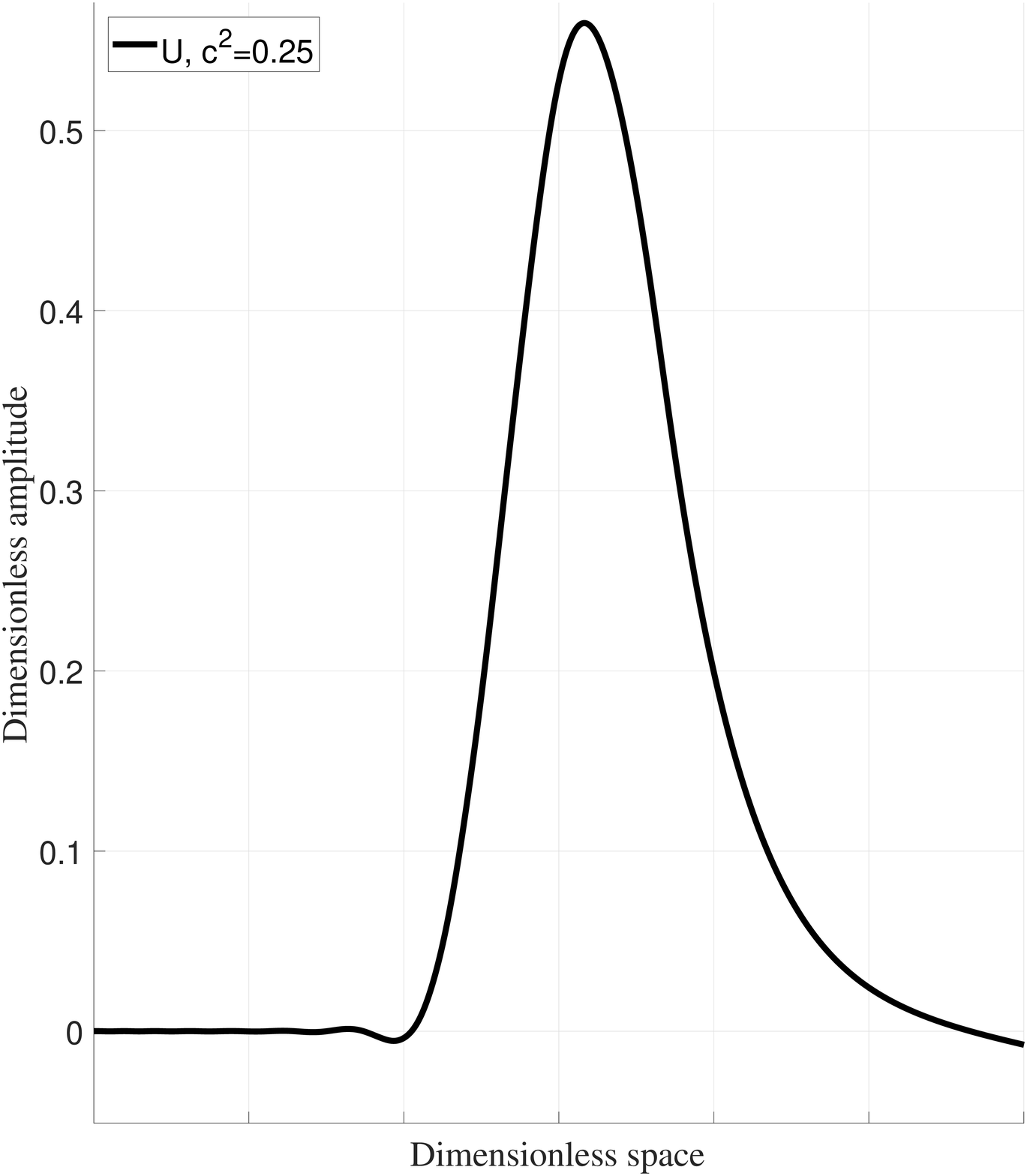}
\includegraphics[width=0.46\textwidth]{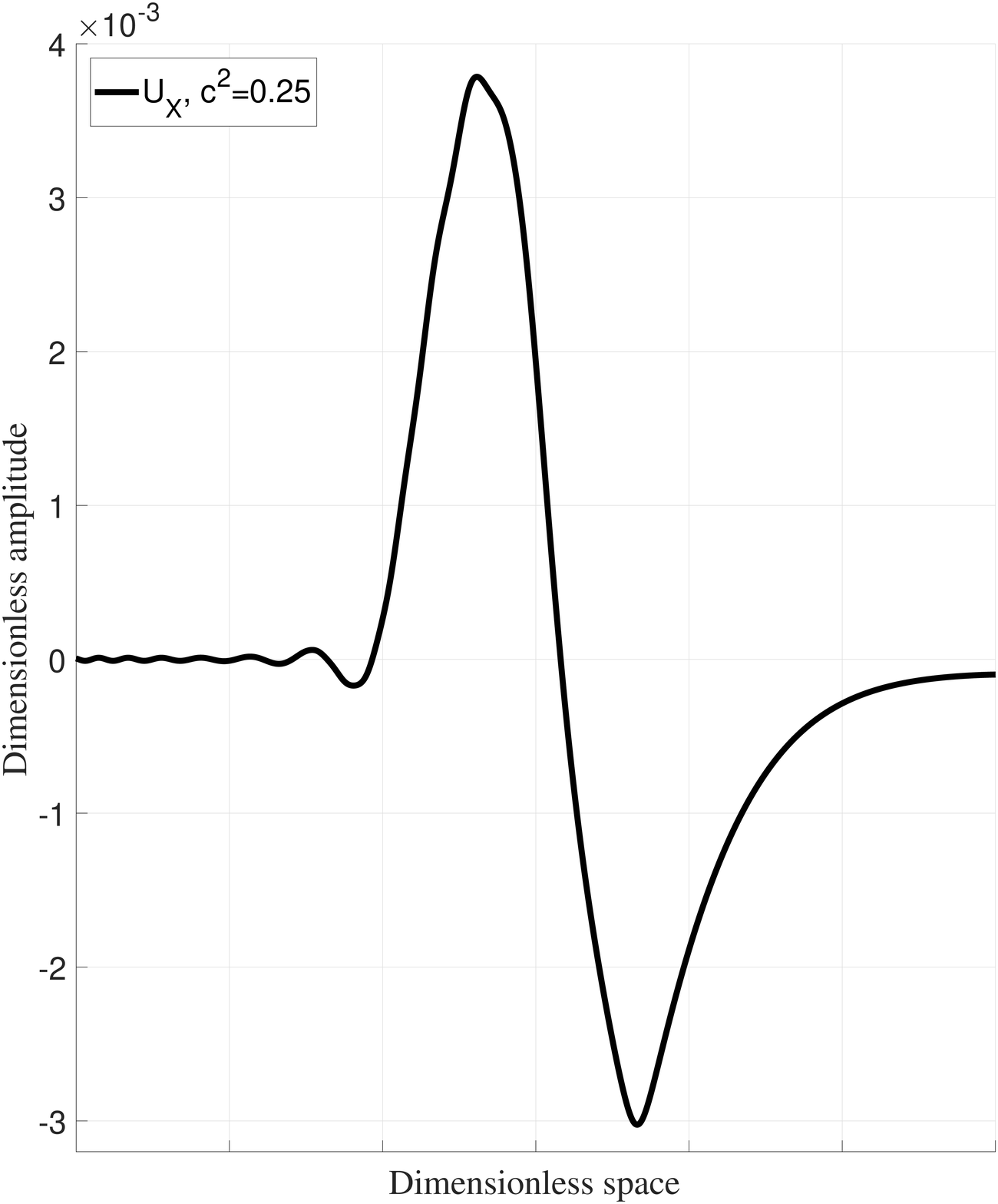}
\caption{Longitudinal density change (left) and transverse displacement (right).} \label{defsiire}
\end{figure}

The numerical results demonstrate a number of characteristics for the combined nerve pulse model which are qualitatively in line with the observations from the experiments. The moving action potential generates a mechanical wave similar to certain extent to results of El Hady and Machta in \cite{Hady2015} although through a different mechanism. It must be stressed that presented profiles are for the longitudinal density change while in experiments usually an transverse displacement is measured. However, the connection between longitudinal density change and transverse displacement can be established in a straightforward manner by drawing inspiration from theory of rods \cite{PeetsTammEngelbrecht2017,Porubov2003}.  The transverse displacement in the theory of rods is proportional to the gradient of the longitudinal density change. In Fig.~\ref{defsiire} a wave-profile of the longitudinal  density change under the considered parameters at $c^2=0.25$ (left) and its gradient (right) are shown. The shape of the gradient is similar to the transverse displacement profiles measured in the experiments \cite{Iwasa1980,Tasaki1988} and as expected, its amplitude is rather small. The shape of transverse displacement derived from the generated mechanical wave (see Fig.~\ref{defsiire}) is in a good agreement with various experimental observations \cite{Iwasa1980,Tasaki1988}.

The numerical results demonstrate that under the proposed model the mechanical wave can influence the characteristics of the electrical signal, like, for example, its shape or propagation velocity. However, under the considered parameters the influence of the mechanical wave is relatively small on the electrical pulse. Then there is a question of velocity synchronization between the different signals in the ensemble. For example, Heimburg and Jackson have measured the unperturbed (low frequency) sound velocity in the lipid layer to be $171.4 \; \mathrm{m/s}$ at $37 \mathrm{C^o}$ and $176.6 \; \mathrm{m/s}$ at $45 \mathrm{C^{o}}$ and the minimum possible velocity for their solitonic solution is assessed to be approximately $115 \; \mathrm{m/s}$ \cite{Heimburg2005} while typically the propagation velocity for the action potential can be much lower than these values. On a related note, in principle, sound velocity in axoplasm which is composed of approximately 80\% of water could be higher than the typical propagation velocity of an action potential as well. However, in the experiment by Terakawa \cite{Terakawa1985} the observed pressure changes were propagating with similar velocities to the propagating action potentials. The numerical results capture the `synchronization' of velocities between mechanical and electrical pulses when tracking the pulse peaks. However, it is also evident that under the considered parameters the wave front of the mechanical wave propagates at higher velocity (the bounding velocity in Fig.~\ref{phase}) under the presented parameter combinations than the action potential. The mechanical effects related to the action potential propagation in axons and lipid bi-layers have been studied in a number of experiments \cite{Fillafer2017,Iwasa1980,Tasaki1988} and the numerical results are qualitatively similar, although, as noted earlier, the results presented here are in the dimensionless form.

The model can also include directly the pressure waves in the axoplasm \cite{Terakawa1985}. In this case the improved Heimburg-Jackson model for the mechanical wave would contain an additional coupling term as a function of the local pressure \cite{Engelbrecht2018d}. The framework can be further extended to include thermal effects \cite{Tamm2019}. 

Finally, the combined model presented here results in a wave ensemble. However, the model has a number of properties
which invite further investigation and analysis. 
The key phenomena observed in the coupled model are: (i) the  velocity of the peak of the mechanical pulse is similar to the velocity  of the action potential regardless of the sound velocity value in the lipid bi-layer, (ii) the velocity of the front and the shape of the mechanical wave depends on sound velocity in the lipid bi-layer, (iii) the shape of the mechanical wave can have an effect on the velocity and shape of the action potential.


\section*{Acknowledgements}
This research was supported by the European Union through the European Regional Development Fund (Estonian
Programme TK 124) and by the Estonian Research Council (projects IUT 33-24, PUT 434). The paper reflects the ideas of a talk at The Tenth IMACS International Conference on
Nonlinear Evolution Equations and Wave Phenomena: Computation and Theory, March 29 -- April 01, 2017. Georgia, USA. 








\newpage
%

\end{document}